# Ionically-Driven Synthesis and Exchange Bias in Mn$_4$N/MnN$_x$ Heterostructures

Zhijie Chen,[1] Christopher J. Jensen,[1] Chen Liu,[2] Xixiang Zhang,[2] and Kai Liu[1,*]

[1]*Physics Department, Georgetown University, Washington, DC 20057, USA*

[2]*King Abdullah University of Science & Technology, Thuwal 23955-6900, Saudi Arabia*

Abstract

Ferrimagnets have received renewed attention as a promising platform for spintronic applications. Of particular interest is the Mn$_4$N from the ε-phase of the manganese nitride as an emergent rare-earth-free spintronic material due to its perpendicular magnetic anisotropy, small saturation magnetization, high thermal stability, and large domain wall velocity. We have achieved high-quality (001)-ordered Mn$_4$N thin film by sputtering Mn onto η-phase Mn$_3$N$_2$ seed layers on Si substrates. As the deposited Mn thickness varies, nitrogen ion migration across the Mn$_3$N$_2$/Mn layers leads to a continuous evolution of the layers to Mn$_3$N$_2$/Mn$_2$N/Mn$_4$N, Mn$_2$N/Mn$_4$N, and eventually Mn$_4$N alone. The ferrimagnetic Mn$_4$N indeed exhibits perpendicular magnetic anisotropy, and forms via a nucleation-and-growth mechanism. The nitrogen ion migration is also manifested in a significant exchange bias, up to 0.3 T at 5 K, due to the interactions between ferrimagnetic Mn$_4$N and antiferromagnetic Mn$_3$N$_2$ and Mn$_2$N. These results demonstrate a promising all-nitride magneto-ionic platform with remarkable tunability for device applications.

*Email: kai.liu@georgetown.edu

Ferrimagnets (FiMs) are characterized by antiparallel-coupled magnetic sublattices with different magnetic moments. They have gained tremendous interests recently as a promising platform for spintronic applications as they offer the combined benefits of ferromagnets (FMs) and antiferromagnets (AFs),[1-4] such as easy manipulation with an external magnetic field and fast spin dynamics.[5] Moreover, some FiMs such as rare-earth iron garnets,[6] rare earth-transition metal films,[1] and $Mn_4N$[7] also possess perpendicular magnetic anisotropy (PMA) under appropriate growth conditions, a desirable functionality for device applications. Among them, $Mn_4N$ stands out due to its high Curie temperature ($T_C$ = 745K),[8] PMA,[7, 9-11] small saturation magnetization (up to 145 emu/cm$^3$),[9] large domain wall velocity,[12] and possibility of hosting non-trivial spin textures.[13, 14] It can also be doped with other elements to change its magnetic properties and compensation point.[15-17] As the only ferrimagnetic phase among the four stable manganese nitrides,[18] namely θ-MnN,[19] η-$Mn_3N_2$,[20, 21] ζ-$Mn_2N$,[22] and ε-$Mn_4N$,[8] $Mn_4N$ has an anti-perovskite structure with two inequivalent and anti-aligned Mn sublattices at the corners and face centers, and a nitrogen atom at the body center.

Furthermore, there have been emerging interests in magneto-ionics, where energy-efficient control of magnetic properties can be achieved via ionic migration.[23-40] Nitrogen-based magneto-ionics has been shown to exhibit good cyclabilities and fast ionic migration.[34, 41-43] Recently, nitride-based exchange bias has been reported in MnN/CoFe heterostructures, which can be manipulated via nitrogen ion migration in and out of the MnN layer using an electric field.[44] This motivates us to build an all-nitride magneto-ionic system taking advantage of the different magnetic phases in Mn nitrides.

$Mn_4N$ thin films are typically grown onto $SrTiO_3$ or MgO substrates at elevated temperatures through molecular beam epitaxy, pulsed laser deposition, or reactive sputtering in a nitrogen environment.[7, 9-12, 14] The film quality is susceptible to the nitrogen flow rate or partial pressure, and the optimum growth conditions vary from study to study.[7, 11] It is challenging to grow high quality thin films of $Mn_4N$ directly on Si substrates,[45, 46] which are CMOS compatible. In this work, we demonstrate that high-quality (001)-ordered $Mn_4N$ thin films can be grown on Si substrates by directly sputtering pure Mn onto an $Mn_3N_2$ seed layer at elevated temperatures, resulted from the chemical reaction between Mn and the nitrogen in the $Mn_3N_2$ seed layer. In a nominally $Mn_3N_2$ (20nm)/Mn ($t_{Mn}$) series of samples, by changing the deposited Mn thickness $t_{Mn}$ from 0 to 50 nm, nitrogen ion migration gradually transforms the layers into $Mn_3N_2/Mn_2N/Mn_4N$, $Mn_2N/Mn_4N$ and eventually $Mn_4N$, confirmed by X-ray diffraction (XRD), transmission electron microscopy (TEM), and electron energy loss spectroscopy (EELS). The $Mn_4N$ films are found to exhibit PMA. First-order reversal curve (FORC) measurements reveal that the $Mn_4N$ forms with a nucleation-and-growth process. The nitrogen ion migration is also manifested in a significant exchange bias, up to 0.3 T at 5 K, due to the interaction between ferrimagnetic $Mn_4N$ and antiferromagnetic $Mn_3N_2$ and $Mn_2N$.

Seed layers of 20 nm $Mn_3N_2$ were first reactive-sputtered onto Si substrate with 285nm thermally oxidized $SiO_2$ layer from a Mn target using direct current (dc) in an ultrahigh vacuum chamber with a base pressure better than $5 \times 10^{-8}$ torr. The substrate temperature was kept at 450 °C, and the Ar : $N_2$ ratio was held at 1 : 1 with a 5 mtorr sputtering gas pressure. These $Mn_3N_2$ films were then left in vacuum for 30 min at the same substrate temperature to promote nitrogen reordering. Subsequently, 0 - 50 nm of Mn was deposited onto the $Mn_3N_2$ layer at the same 450 °C substrate temperature in an Ar-only environment. After deposition, substrate heating was immediately turned off, and the samples were cooled to room temperature before depositing a 5 nm Ti capping layer.

Structural characterizations were performed using XRD on a Panalytical X'Pert³ MRD with symmetric 2θ-ω and grazing incidence geometries. Sample microstructures and composition analysis were done using an FEI Titan Themes Cubed G2 300 (Cs Probe) TEM at KAUST.

Magnetic measurements were carried out using a Quantum Design superconducting quantum interference device (SQUID) magnetometer. Exchange bias was measured at 5 K by first field-cooling the sample from 300 K in a 2 T magnetic field, all in the out-of-plane (OP) geometry. FORC measurements[47-51] were done in a vibrating sample magnetometer at room temperature by saturating samples in a 1.5 T OP magnetic field and then measuring from a reversal field ($H_R$) back to saturation. This process was repeated at different $H_R$ to fill the interior of the hysteresis loop, creating a family of FORCs. A FORC distribution was then calculated using Eqn. (1),

$$\rho\,(H, H_R) \equiv -\frac{1}{2M_S}\frac{\partial^2 M(H, H_R)}{\partial H\, \partial H_R}. \qquad (1)$$

where $M_S$ is the saturation magnetization, and $M(H, H_R)$ is the magnetization at the applied field $H$ with reversal field $H_R$. The FORC distribution can also be represented in terms of local coercive field and bias field $(H_C, H_B)$ defined by $H_C = \frac{1}{2}(H - H_R)$ and $H_B = \frac{1}{2}(H + H_R)$.

The η-phase $Mn_3N_2$ is chosen as the seed layer for $Mn_4N$ growth because it provides the crystalline texture and nitrogen needed for the $Mn_4N$ growth.[52] As shown in Fig. 1(a) and (b), $Mn_3N_2$ is antiferromagnetic ($T_N$ ~925 K), and its c-axis is about three times that of $Mn_4N$.[21, 53] XRD reveals the growth of the $Mn_3N_2$ phase, shown in Fig. 1(c), with a preferred orientation along the (010) direction and the c-axis in the film plane. Furthermore, grazing incidence XRD confirms that all peaks are from $Mn_3N_2$ (Supplementary Material Fig. S1). Upon depositing 40 nm of Mn, XRD shows that the film is primarily the $Mn_4N$ phase oriented along (001), with no appreciable $Mn_3N_2$ phase left, as shown in Fig. 1(d) and Supplementary Material Fig. S1. Along with the structure changes, there is also a drastic change in the film magnetic properties. As shown in Fig. 1(e), the initial $Mn_3N_2$ layer does not exhibit any magnetic signal, consistent with its antiferromagnetic nature; interestingly, the sample deposited with Mn exhibits a square loop with a large coercivity (0.27 T) and a small $M_S$ (85 emu/cm$^3$) that are typical of $Mn_4N$ films.[7, 9, 10]

To understand how the film transforms from $Mn_3N_2$ to $Mn_4N$ by only depositing Mn, we have investigated a series of samples starting with 20 nm $Mn_3N_2$ seed layer, but the deposited Mn nominal thickness varied from 0 to 50 nm with a 5 nm step size. From now on, each sample is referred by its deposited Mn thickness ($t_{Mn}$) unless otherwise stated. Fig. 2(a) reveals how the Mn nitride phase evolves across the samples. Starting from $t_{Mn}$ = 0 nm, which is the $Mn_3N_2$ layer, the only peak is the $Mn_3N_2$ (020). As $t_{Mn}$ increases, a prominent peak emerges around 47.2°, corresponding to the $Mn_4N$ (002), indicating $Mn_4N$ formation as Mn is deposited. On the other hand, the $Mn_3N_2$ (020) peak diminishes and shifts to higher angles before eventually vanishing in the $t_{Mn}$ = 30 nm sample. This trend indicates that the $Mn_3N_2$ phase is fading and is not as stable as $Mn_4N$ at high temperatures, consistent with prior studies.[19, 54] Interestingly, as the $Mn_3N_2$ peak gets smaller, another peak emerges near 42.2° in the $t_{Mn}$ = 20 nm sample and grows larger before disappearing in the $t_{Mn}$ = 40 nm sample. This peak is the (111) diffraction from the ζ-phase $Mn_2N_{0.86}$, which has a thermal stability and nitrogen content between η-$Mn_3N_2$ and ε-$Mn_4N$.[18, 19, 54] In the $t_{Mn}$ = 40 and 45 nm samples, both $Mn_3N_2$ and $Mn_2N$ peaks have vanished, while the $Mn_4N$ (002) peak gets even larger and closer to its expected location. Eventually, when $t_{Mn}$ reaches 50 nm, the α-Mn (221) peak shows up near 43°, indicating that some deposited Mn remain unreacted as the entire nitride film is now $Mn_4N$. Grazing incidence scans are included in the Supplementary Material Fig. S1, consistent with Fig. 2(a), along with full range 2θ-ω scans (Fig. S2).

The Mn$_4$N crystallite size has been estimated from the full-width-at-half-maximum (FWHM) of the (002) peak, after instrument broadening correction, using the Scherrer equation.[55] The Mn$_4$N crystallite size nearly doubles as more Mn is deposited, reaching a plateau after $t_{Mn}$ = 40 nm, as shown in Fig. 2(b). This is consistent with the trend in Fig. 2(a), where the Mn$_4$N peak becomes sharper and more prominent, and Mn$_4$N is the only phase after $t_{Mn}$ reaches 40 nm. This crystallite size estimation is rather simplified, as it ignores peak width contribution from other factors such as inhomogeneities in $d$-spacing. As the film stoichiometry changes due to the nitrogen migration, any spread in N-content and the lattice parameters would lead to a broadening of the peak width. Interestingly, the overall narrowing trend of the Mn$_4$N peak width with increasing $t_{Mn}$ suggests that the stoichiometry variation is suppressed at high $t_{Mn}$, which is consistent with the fact that when $t_{Mn}$ reaches 40 nm, only a single phase Mn$_4$N is observed. Moreover, the peak locations shift to higher angles as $t_{Mn}$ increases, as shown in Fig 2(c). Nitride phases' lattice constants are known to be very sensitive to nitrogen content,[18] as interstitial nitrogen usually causes the lattices to expand. As Mn$_3$N$_2$ loses nitrogen to the deposited Mn, its lattice contracts, causing the Mn$_3$N$_2$ peak to shift to a higher angle until this phase is gone. The Mn$_4$N peak location, on the other hand, stays relatively constant before changing rapidly beyond $t_{Mn}$ = 40 nm, likely caused by the nitrogen redistribution within the Mn$_4$N phase once the nitrogen from Mn$_3$N$_2$ and Mn$_2$N has been depleted. Thus, we postulate that as Mn is deposited onto the Mn$_3$N$_2$ layer at elevated temperatures, it reacts with the nitrogen coming from Mn$_3$N$_2$ and forms Mn$_4$N. While Mn$_3$N$_2$ loses nitrogen, it first turns into Mn$_2$N and eventually becomes Mn$_4$N. These reactions are summarized in Eqn. 2a, and they can be combined into one chemical reaction (Eqn. 2b) since they are multistep reactions.

$$2Mn_3N_2 \rightarrow 3Mn_2N + N; \quad 4Mn + N \rightarrow Mn_4N; \quad 2Mn_2N \rightarrow Mn_4N + N \qquad (2a)$$

$$5Mn + Mn_3N_2 \rightarrow 2Mn_4N; \qquad (2b)$$

We also calculated the enthalpy of formation for the reaction shown in Eqn. 2b to be -110 kJ/mol using the standard enthalpy of formation for $Mn_4N$ and $Mn_3N_2$,[54] indicating that this reaction is thermodynamically favorable.

Cross-sectional TEM and EELS lines scans were performed to explore the microstructure and nitrogen content evolution as $t_{Mn}$ varies. Fig. 2(d) shows the EELS line scan along the film thickness (green line in the inset) for $t_{Mn}$ = 0 nm. The Mn:N composition ratio is determined to be 58:42, consistent with the nominal atomic ratio of $Mn_3N_2$. As $t_{Mn}$ increases to 40 nm, the EELS scan shown in Fig. 2(e) illustrates that the Mn:N ratio is now 84:16, also consistent with the nominal atomic ratio of $Mn_4N$. Interestingly, most of the nitride layers appear homogenous with constant Mn:N ratio from both the cross-sectional TEM and EELS, indicating that nitrogen in the $Mn_3N_2$ seed layer has redistributed to maintain a constant nitrogen concentration within the Mn nitrides after the Mn is deposited. These results further corroborate our postulation that nitrogen moves from the $Mn_3N_2$ seed layer into the Mn layer to form more stable $Mn_4N$.

We then investigate the magnetic properties of this series of samples. Fig. 3(a) shows the room temperature hysteresis loops with in-plane (IP) and out-of-plane (OP) magnetic fields. The OP loops get more square and broader as $t_{Mn}$ increases from 10 to 50 nm while the IP loops stay relatively constant. These trends are further revealed by plotting the squareness, or ratio of remanence magnetization ($M_r$) over $M_S$, for each sample, Fig. 3 (b). The OP and IP remanence are small and stay relatively constant for $t_{Mn}$ < 20 nm, indicating the lack of a clear magnetic easy axis. For 20 nm < $t_{Mn}$ < 35 nm, a sharp jump in OP remanence is observed, along with a drop in IP remanence, indicating a clear easy axis has been established in the OP direction. At $t_{Mn} > 35$ nm, the easy axis remains OP, while IP remanence increases slightly but remains low.

We have also calculated the uniaxial magnetic anisotropy constant ($K_u$, see Supplementary Material). As shown in Fig. 3(c), $K_u$ starts out to be negative for $t_{Mn} = 5$ nm and shows a clear switching from negative to positive, especially when $t_{Mn} > 20$ nm, further confirming the magnetic easy axis switching to OP as more Mn₄N is formed. Note that $K_u$ exhibits the largest value (0.03 MJ/m³), or the film has the largest PMA when 35 nm < $t_{Mn}$ < 45 nm. This is also consistent with the XRD result, which shows that Mn₄N is the only phase for this $t_{Mn}$ range. This $K_u$ value is smaller than other reported values which range from 0.05 to 0.2 MJ/m³.[7, 9-11, 56, 57] The uniaxial anisotropy has been attributed to the tetragonal distortion caused by in-plane tensile strains.[7, 9, 10] However, in this study, the films that show the largest PMA has *c/a* close to 1 (Table 1 in Supplementary Material). The lack of in-plane strain may be the reason for our smaller $K_u$ values since the films are deposited onto an amorphous SiO₂ layer. Additionally, the sizeable PMA in our films may have come from other contributions such as the shape anisotropy of the Mn₄N grains.[58]

To investigate how the Mn₄N phase evolves with $t_{Mn}$ and the corresponding magnetization reversal, we have carried out FORC studies in the OP geometry at room temperature, as shown in Fig. 4. For the $t_{Mn} = 10$ and 20 nm samples, individual FORCs fill the major loops in a slanted fashion, Fig. 4a and 4c, respectively. The corresponding FORC distributions exhibit a prominent vertical ridge centered around $H_C = 0$, which corresponds to reversible switching,[51, 59] and a smaller horizontal feature centered at $\mu_0 H_C = 120$ mT and 150 mT, respectively (Fig. 4b, 4d). This indicates that the Mn₄N film is mainly reversible and magnetically soft. Likely for this $t_{Mn}$ range, the Mn₄N phase is just emerging in small clusters scattered in an antiferromagnetic matrix of Mn$_3$N$_2$ and Mn$_2$N. As more Mn$_4$N is formed, families of FORCs for $t_{Mn}$ = 30, 40, and 50 nm are considerably different, as individual FORCs return to positive saturation in a more horizontal fashion, consistent with the establishment of a magnetic easy axis (Figs. 4e, 4g, 4i). Their FORC distributions are also strikingly different. The previous large vertical reversible ridge at $\mu_0 H_C = 0$ becomes smaller and eventually vanishes in the $t_{Mn}$ = 50 nm sample. The horizontal feature along the $H_C$ axis now becomes prominent and shifts to higher $\mu_0 H_C$ of 460, 310, and 390 mT, respectively (Fig. 4f, 4h, 4j). The change in relative intensity of the horizontal and vertical FORC features likely indicate that Mn$_4$N forms via a nucleation-and-growth mechanism, similar to that reported previously in the ordering of $L1_0$ FeCuPt.[59]

In this nominally Mn$_3$N$_2$ (20nm) /Mn ($t_{Mn}$) series of samples, the evolution of the AF phase and the emergence of the FiM phase are also manifested in the exchange bias effect, which was studied at 5 K after cooling the samples from room temperature in a positive 2 T OP magnetic field. A significant horizontal shift to the negative field direction, up to 300 mT, and a coercivity enhancement can be seen, Fig. 5(a), typical of exchange bias systems.[44, 60-62] The $t_{Mn}$ dependence of coercivity ($H_C$) and exchange field ($H_E$) both exhibit non-monotonic trends, with an intriguing peak around 20 nm < $t_{Mn}$ < 30 nm, as shown in Fig. 5(b). These trends are likely a combined effect from the AF phase evolution as well as the FiM thickness and $M_S$ variations. To further explore the exchange anisotropy independent of the FiM, we have evaluated the interfacial exchange energy ($J_{int}$) per unit area using Eqn. 3:[60, 62]

$$J_{int} = M_{FiM} t_{FiM} H_E = m_{FiM} H_E / A \tag{3}$$

where $M_{FiM}$, $m_{FiM}$, and $t_{FiM}$ are the FiM saturation magnetization, saturation magnetic moment, and layer thickness, respectively, $H_E$ is the exchange field, and $A$ is the sample area. As shown in Fig. 5(c), the dependence of $J_{int}$ on $t_{Mn}$ exhibits a bell-shaped plot that peaks around 20 to 30 nm. $J_{int}$ is small and increases continuously for $t_{Mn}$ of 5 - 15 nm, where the dominating AF phase is Mn$_3$N$_2$, as observed by XRD. Due to the high T$_N$ of Mn$_3$N$_2$, only a small fraction of the AF spins is aligned to pin Mn$_4$N by field cooling from room temperature, resulting in a small $J_{int}$. However, for 20 nm < $t_{Mn}$ <35 nm samples, another AF phase, Mn$_2$N, starts to dominate. By field cooling from 300 K, the Mn$_2$N is effectively coupled with Mn$_4$N, resulting in a significant exchange bias at 5 K. Exchange energy then quickly decreases as Mn$_2$N is turned into Mn$_4$N. By $t_{Mn}$ = 40 nm, no AF phases can be identified from XRD and $J_{int}$ mostly vanishes.

In summary, we have achieved high-quality Mn$_4$N films growth by depositing pure Mn onto an Mn$_3$N$_2$ seed layer. By varying the Mn thickness $t_{Mn}$, the nitrogen concentration in the starting Mn$_3$N$_2$/Mn bilayers can be continuously tuned to be Mn$_3$N$_2$/Mn$_2$N/Mn$_4$N, Mn$_2$N/Mn$_4$N, and eventually to Mn$_4$N alone, as observed by XRD and TEM/EELS. With increasing $t_{Mn}$, more Mn$_4$N is formed, with an increasing PMA reaching 0.03 MJ/m$^3$. FORC measurements further reveal that Mn$_4$N forms via a nucleation-and-growth mechanism. A large exchange bias up to 0.3 T is found at 5 K in this all-nitride system. The variation of the exchange anisotropy is further attributed to the phase change of the antiferromagnets caused by nitrogen redistribution. These results demonstrate an effective all-nitride magneto-ionic platform for studying the properties of the emergent ferrimagnetic Mn$_4$N and its potential applications in spintronics.

**Supplementary Material**

Grazing incidence and full range 2θ-ω X-ray diffraction patterns, Mn nitride phase evolution with layer thickness, cross-sectional TEM and EELS line scans, uniaxial magnetic anisotropy calculation, and additional sample information (PDF).

**Acknowledgement**

This work has been supported in part by the NSF (DMR-2005108, ECCS-2151809), by SMART (2018-NE-2861), one of seven centers of nCORE, a Semiconductor Research Corporation program, sponsored by National Institute of Standards and Technology (NIST), and by KAUST (OSR-2019-CRG8-4081). The acquisition of a Magnetic Property Measurements System (MPMS3), which was used in this investigation was supported by the NSF-MRI program (DMR-1828420).

**Figures**

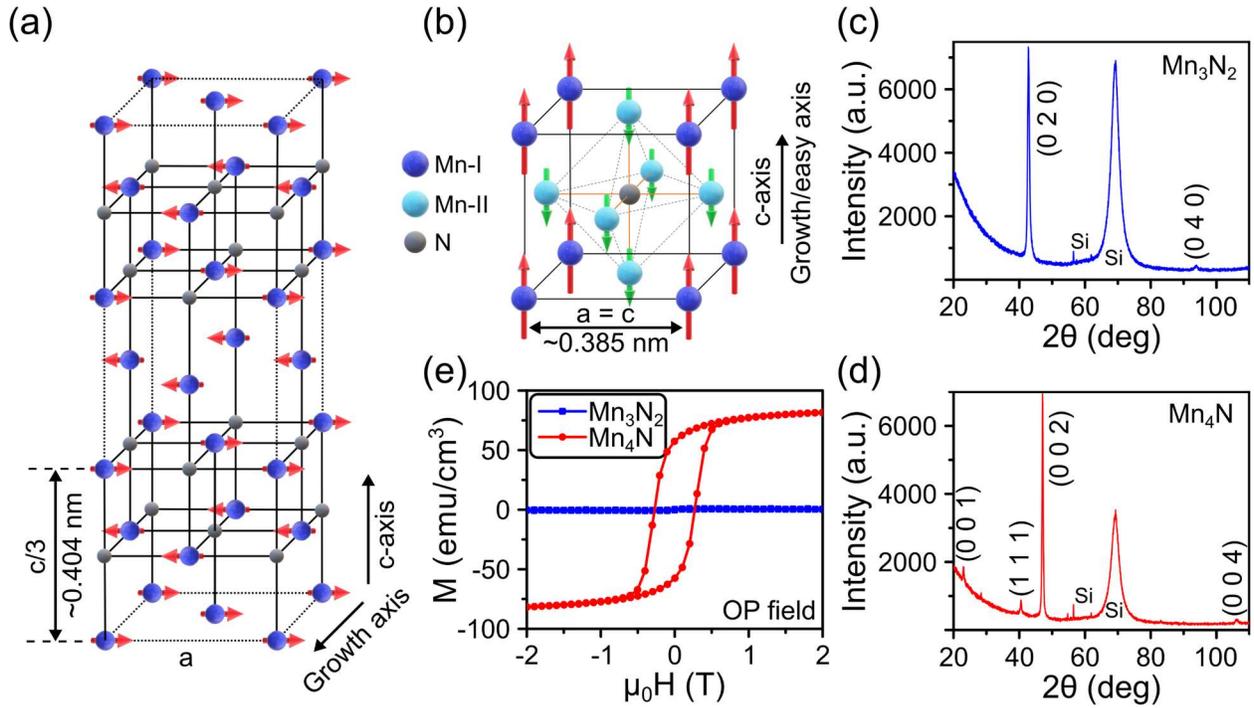

**Figure 1**. Schematics showing the lattice structure and spin orientation of (a) $Mn_3N_2$ and (b) $Mn_4N$. XRD 2θ-ω scans of (c) 20 nm $Mn_3N_2$ seed layer and (d) $Mn_4N$ sample fabricated by depositing 40 nm Mn on top of 20 nm $Mn_3N_2$. (e) Room temperature out-of-plane hysteresis loops for the same $Mn_3N_2$ and $Mn_4N$ samples.

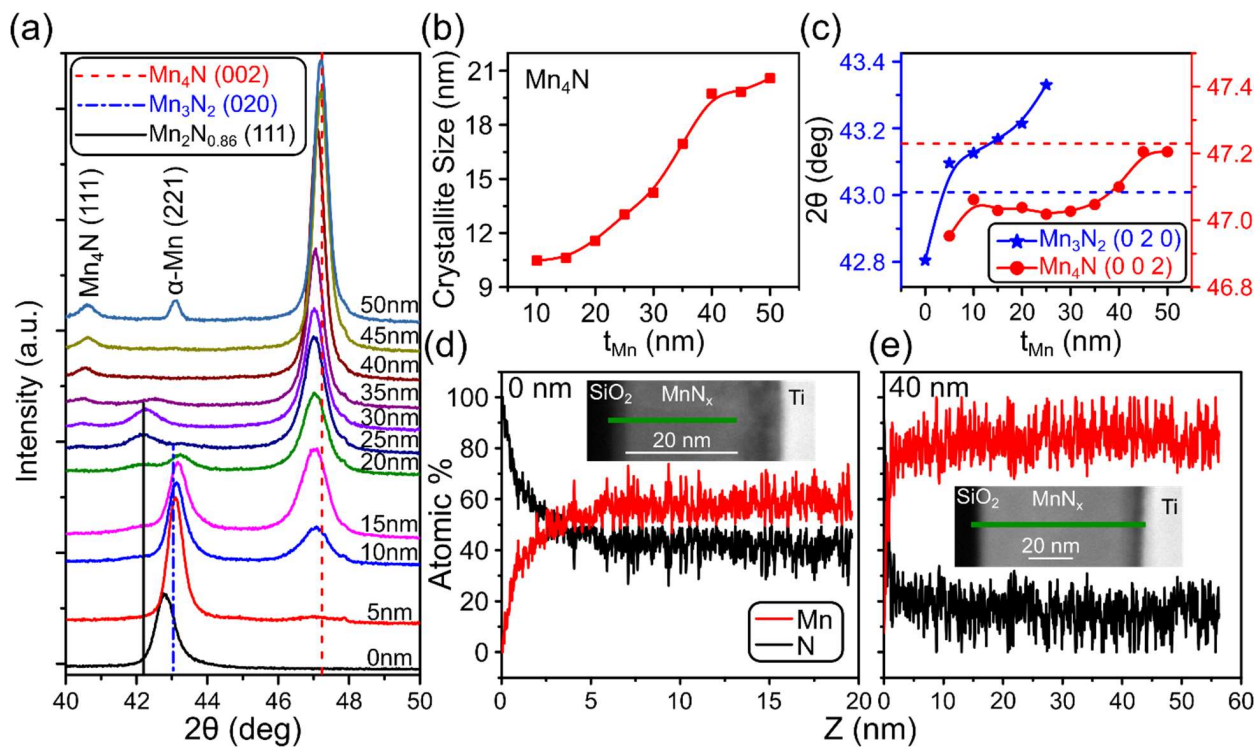

**Figure 2**. (a) XRD 2θ-ω scans for nominally $Mn_3N_2$ (20nm) / Mn ($t_{Mn}$) films, $t_{Mn}$ is the deposited Mn thickness listed next to each curve. Vertical lines show the expected peak locations of $Mn_4N$ (002) (red), $Mn_3N_2$ (020) (blue), and $Mn_2N_{0.86}$ (111) (black). Trends showing (b) $Mn_4N$ crystallite size and (c) $Mn_4N$ (002) (red) and $Mn_3N_2$ (020) (blue) peak location variations. Solid lines are guides to the eye. Dotted horizontal lines are the expected peak locations for $Mn_4N$ (002) (red) and $Mn_3N_2$ (020) (blue). EELS scans along the green line in the TEM insets showing Mn:N ratio across the sample for (d) $t_{Mn}$ = 0 nm and (e) $t_{Mn}$ = 40 nm, where $Z$ = 0 is the starting point of the interface between the substrate and $MnN_x$ layers.

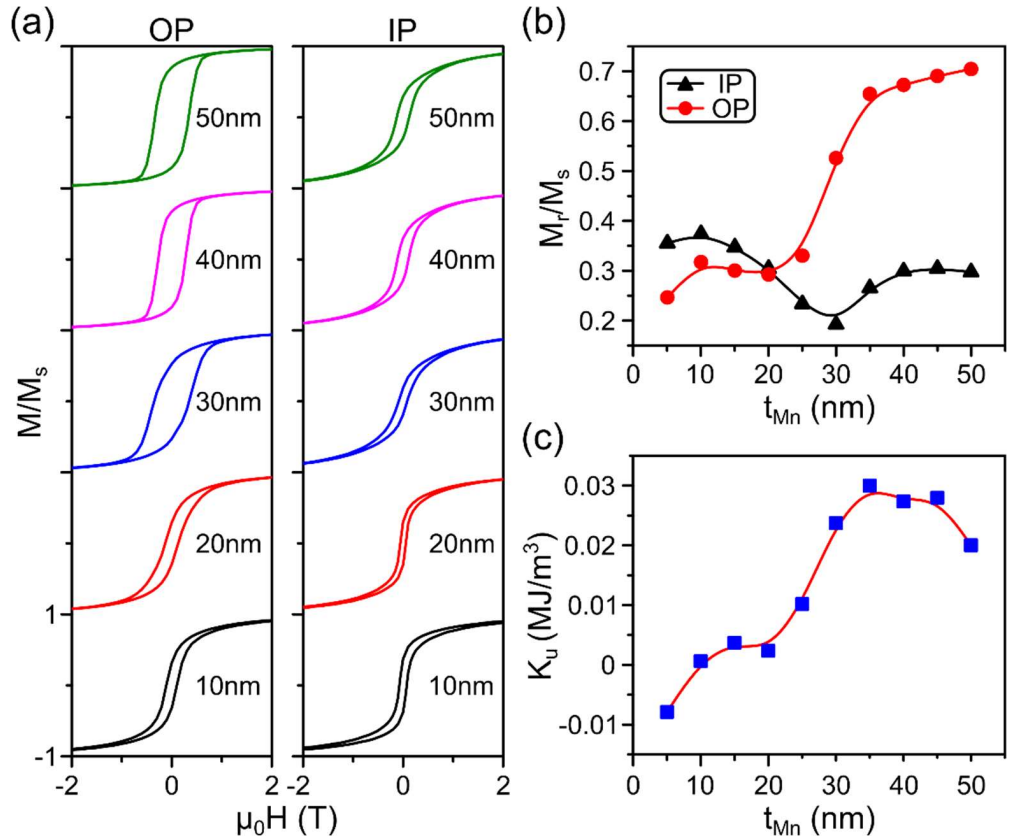

**Figure 3.** (a) Room temperature hysteresis loop with out-of-plane (left) and in-plane (right) magnetic fields for nominally Mn$_3$N$_2$ (20nm) / Mn ($t_{Mn}$) films, where $t_{Mn}$ is the deposited Mn thickness listed next to each curve. Trends for the (b) in-plane and out-of-plane remanence and (c) $K_u$ as $t_{Mn}$ increases. Solid lines in (b) and (c) are guides to the eye.

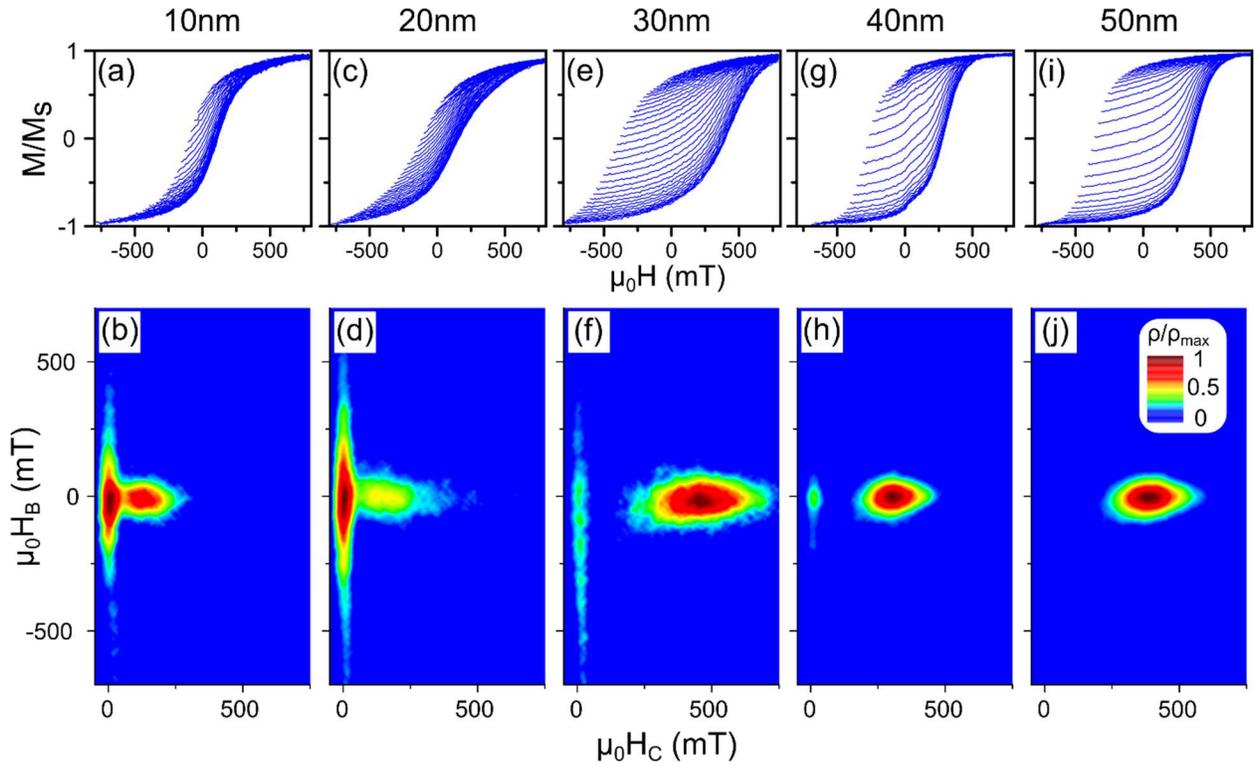

**Figure 4**. Families of FORCs (top row) and FORC distributions (bottom row) for (a,b) $t_{Mn}$ = 10 nm, (c,d) 20 nm, (e,f) 30 nm, (g,h) 40 nm, and (i,j) 50 nm. $t_{Mn}$ is the deposited Mn thickness on top of 20 nm $Mn_3N_2$ seed layer.

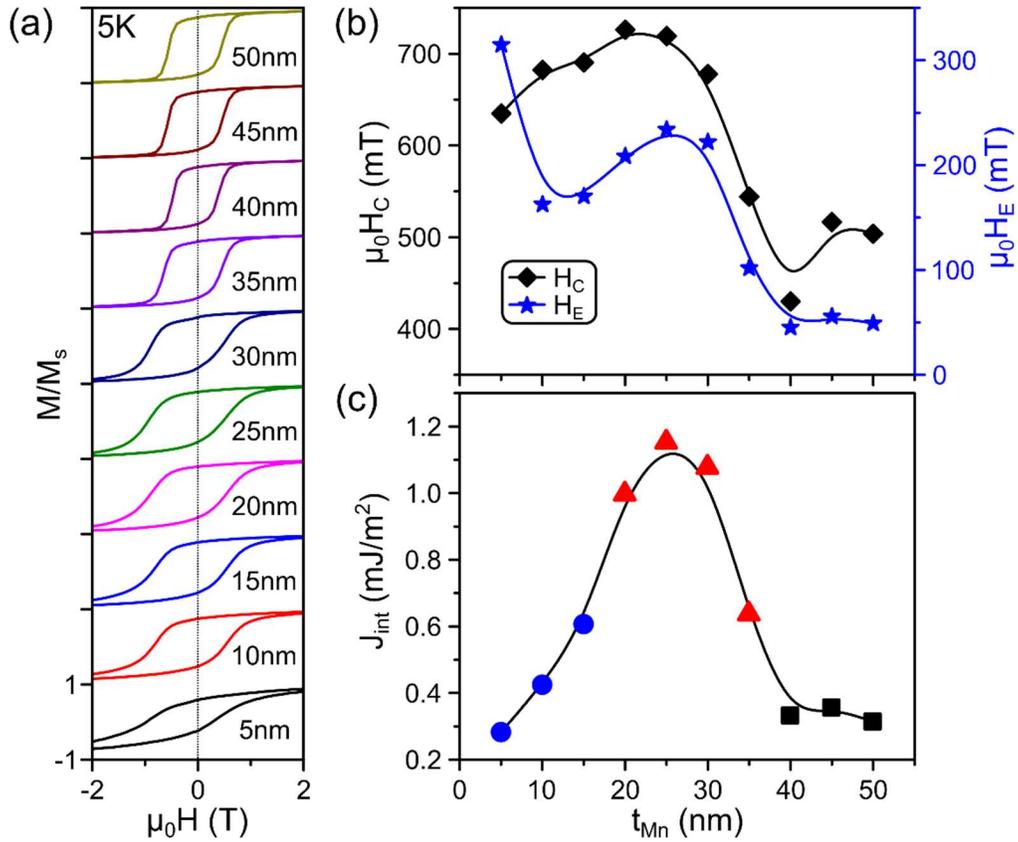

**Figure 5**. (a) Hysteresis loops of the nominally $Mn_3N_2$ (20 nm)/Mn ($t_{Mn}$) series of samples measured at 5 K after 2 T field-cooling from 300 K, with the $t_{Mn}$ listed for each sample. (b) Dependence of coercivity (black) and exchange field (blue) on $t_{Mn}$ at 5 K. (c) Dependence of interfacial exchange energy on $t_{Mn}$ at 5 K where the color represents the different dominating antiferromagnetic phases: $Mn_3N_2$ (blue), $Mn_2N$ (red), and no antiferromagnets (black). Solid lines in (b) and (c) are guides to the eye.

# References

1. L. Caretta, M. Mann, F. Buttner, K. Ueda, B. Pfau, C. M. Gunther, P. Hessing, A. Churikova, C. Klose, M. Schneider, D. Engel, C. Marcus, D. Bono, K. Bagschik, S. Eisebitt, and G. S. D. Beach, Nat. Nanotechnol. **13**, 1154 (2018).

2. R. Blasing, T. P. Ma, S. H. Yang, C. Garg, F. K. Dejene, A. T. N'Diaye, G. Chen, K. Liu, and S. S. P. Parkin, Nat. Commun. **9**, 4984 (2018).

3. J. Finley and L. Liu, Appl. Phys. Lett. **116**, 110501 (2020).

4. S. K. Kim, G. S. D. Beach, K.-J. Lee, T. Ono, T. Rasing, and H. Yang, Nat. Mater. **21**, 24 (2022).

5. V. Baltz, A. Manchon, M. Tsoi, T. Moriyama, T. Ono, and Y. Tserkovnyak, Rev. Mod. Phys. **90**, 015005 (2018).

6. C. O. Avci, A. Quindeau, C. F. Pai, M. Mann, L. Caretta, A. S. Tang, M. C. Onbasli, C. A. Ross, and G. S. Beach, Nat. Mater. **16**, 309 (2017).

7. K. Kabara and M. Tsunoda, J. Appl. Phys. **117**, 17b512 (2015).

8. W. J. Takei, R. R. Heikes, and G. Shirane, Phys. Rev. **125**, 1893 (1962).

9. Y. Yasutomi, K. Ito, T. Sanai, K. Toko, and T. Suemasu, J. Appl. Phys. **115**, 17a935 (2014).

10. T. Hirose, T. Komori, T. Gushi, A. Anzai, K. Toko, and T. Suemasu, AIP Adv. **10**, 025117 (2020).

11. S. Isogami, K. Masuda, and Y. Miura, Phys. Rev. Materials **4**, 014406 (2020).

12. T. Gushi, M. Jovicevic Klug, J. Pena Garcia, S. Ghosh, J. P. Attane, H. Okuno, O. Fruchart, J. Vogel, T. Suemasu, S. Pizzini, and L. Vila, Nano Lett. **19**, 8716 (2019).

13. T. Bayaraa, C. Xu, and L. Bellaiche, Phys. Rev. Lett. **127**, 217204 (2021).

14. C. T. Ma, T. Q. Hartnett, W. Zhou, P. V. Balachandran, and S. J. Poon, Appl. Phys. Lett. **119**, 192406 (2021).

15. K. Ito, Y. Yasutomi, S. Zhu, M. Nurmamat, M. Tahara, K. Toko, R. Akiyama, Y. Takeda, Y. Saitoh, T. Oguchi, A. Kimura, and T. Suemasu, Phys. Rev. B **101**, 104401 (2020).

16. S. Ghosh, T. Komori, A. Hallal, J. Pena Garcia, T. Gushi, T. Hirose, H. Mitarai, H. Okuno, J. Vogel, M. Chshiev, J. P. Attane, L. Vila, T. Suemasu, and S. Pizzini, Nano Lett. **21**, 2580 (2021).

[17] S. Isogami, M. Ohtake, and Y. K. Takahashi, J. Appl. Phys. **131**, 073904 (2022).

[18] N. A. Gokcen, Bull. Alloy Phase Diagrams **11**, 33 (1990).

[19] K. Suzuki, T. Kaneko, H. Yoshida, Y. Obi, H. Fujimori, and H. Morita, J. Alloys Compd. **306**, 66 (2000).

[20] A. Leineweber, R. Niewa, H. Jacobs, and W. Kockelmann, J. Mater. Chem. **10**, 2827 (2000).

[21] H. Yang, H. Al-Brithen, A. R. Smith, J. A. Borchers, R. L. Cappelletti, and M. D. Vaudin, Appl. Phys. Lett. **78**, 3860 (2001).

[22] Y. Liu, L. Xu, X. Li, P. Hu, and S. Li, J. Appl. Phys. **107**, 103914 (2010).

[23] M. Weisheit, S. Fähler, A. Marty, Y. Souche, C. Poinsignon, and D. Givord, Science **315**, 349 (2007).

[24] U. Bauer, L. Yao, A. J. Tan, P. Agrawal, S. Emori, H. L. Tuller, S. van Dijken, and G. S. D. Beach, Nat. Mater. **14**, 174 (2015).

[25] C. Bi, Y. Liu, T. Newhouse-Illige, M. Xu, M. Rosales, J. W. Freeland, O. Mryasov, S. Zhang, S. G. E. te Velthuis, and W. G. Wang, Phys. Rev. Lett. **113**, 267202 (2014).

[26] D. A. Gilbert, J. Olamit, R. K. Dumas, B. J. Kirby, A. J. Grutter, B. B. Maranville, E. Arenholz, J. A. Borchers, and K. Liu, Nat. Commun. **7**, 11050 (2016).

[27] D. A. Gilbert, A. J. Grutter, E. Arenholz, K. Liu, B. J. Kirby, J. A. Borchers, and B. B. Maranville, Nat. Commun. **7**, 12264 (2016).

[28] K. Duschek, D. Pohl, S. Fähler, K. Nielsch, and K. Leistner, APL Mater. **4**, 032301 (2016).

[29] J. Walter, H. Wang, B. Luo, C. D. Frisbie, and C. Leighton, ACS Nano **10**, 7799 (2016).

[30] A. Quintana, E. Menendez, M. O. Liedke, M. Butterling, A. Wagner, V. Sireus, P. Torruella, S. Estrade, F. Peiro, J. Dendooven, C. Detavernier, P. D. Murray, D. A. Gilbert, K. Liu, E. Pellicer, J. Nogues, and J. Sort, ACS Nano **12**, 10291 (2018).

[31] A. J. Tan, M. Huang, C. O. Avci, F. Büttner, M. Mann, W. Hu, C. Mazzoli, S. Wilkins, H. L. Tuller, and G. S. D. Beach, Nat. Mater. **18**, 35 (2019).

[32] L. Herrera Diez, Y. T. Liu, D. A. Gilbert, M. Belmeguenai, J. Vogel, S. Pizzini, E. Martinez, A. Lamperti, J. B. Mohammedi, A. Laborieux, Y. Roussigné, A. J. Grutter, E. Arenholtz, P. Quarterman,

B. Maranville, S. Ono, M. S. E. Hadri, R. Tolley, E. E. Fullerton, L. Sanchez-Tejerina, A. Stashkevich, S. M. Chérif, A. D. Kent, D. Querlioz, J. Langer, B. Ocker, and D. Ravelosona, Phys. Rev. Appl. **12**, 034005 (2019).

33 J. Zehner, R. Huhnstock, S. Oswald, U. Wolff, I. Soldatov, A. Ehresmann, K. Nielsch, D. Holzinger, and K. Leistner, Adv. Electron. Mater. **5**, 1900296 (2019).

34 J. de Rojas, A. Quintana, A. Lopeandía, J. Salguero, B. Muñiz, F. Ibrahim, M. Chshiev, A. Nicolenco, M. O. Liedke, M. Butterling, A. Wagner, V. Sireus, L. Abad, C. J. Jensen, K. Liu, J. Nogués, J. L. Costa-Krämer, E. Menéndez, and J. Sort, Nat. Commun. **11**, 5871 (2020).

35 G. Chen, A. Mascaraque, H. Jia, B. Zimmermann, M. Robertson, R. L. Conte, M. Hoffmann, M. A. González Barrio, H. Ding, R. Wiesendanger, E. G. Michel, S. Blügel, A. K. Schmid, and K. Liu, Sci. Adv. **6**, eaba4924 (2020).

36 J. Walter, B. Voigt, E. Day-Roberts, K. Heltemes, R. M. Fernandes, T. Birol, and C. Leighton, Sci. Adv. **6**, eabb7721 (2020).

37 G. Chen, M. Robertson, M. Hoffmann, C. Ophus, A. L. F. Cauduro, R. Lo Conte, H. F. Ding, R. Wiesendanger, S. Blügel, A. K. Schmid, and K. Liu, Phys. Rev. X **11**, 021015 (2021).

38 Y. Guan, X. Zhou, F. Li, T. Ma, S.-H. Yang, and S. S. P. Parkin, Nat. Commun. **12**, 5002 (2021).

39 Q. Wang, Y. Gu, C. Chen, F. Pan, and C. Song, J. Phys. Chem. Lett. **13**, 10065 (2022).

40 G. Chen, C. Ophus, A. Quintana, H. Kwon, C. Won, H. Ding, Y. Wu, A. K. Schmid, and K. Liu, Nat. Commun. **13**, 1350 (2022).

41 J. de Rojas, J. Salguero, F. Ibrahim, M. Chshiev, A. Quintana, A. Lopeandia, M. O. Liedke, M. Butterling, E. Hirschmann, A. Wagner, L. Abad, J. L. Costa-Kramer, E. Menendez, and J. Sort, ACS Appl. Mater. Interfaces **13**, 30826 (2021).

42 J. de Rojas, A. Quintana, G. Rius, C. Stefani, N. Domingo, J. L. Costa-Krämer, E. Menéndez, and J. Sort, Appl. Phys. Lett. **120**, 070501 (2022).

43 Z. Tan, S. Martins, M. Escobar, J. de Rojas, F. Ibrahim, M. Chshiev, A. Quintana, A. Lopeandia, J. L. Costa-Krämer, E. Menéndez, and J. Sort, ACS Appl. Mater. Interfaces **14**, 44581 (2022).

44  C. J. Jensen, A. Quintana, P. Quarterman, A. J. Grutter, P. P. Balakrishnan, H. Zhang, A. V. Davydov, X. Zhang, and K. Liu, ACS Nano **17**, 6745 (2023).

45  K. M. Ching, W. D. Chang, T. S. Chin, J. G. Duh, and H. C. Ku, J. Appl. Phys. **76**, 6582 (1994).

46  S. Nakagawa and M. Naoe, J. Appl. Phys. **75**, 6568 (1994).

47  C. R. Pike, A. P. Roberts, and K. L. Verosub, J. Appl. Phys. **85**, 6660 (1999).

48  J. E. Davies, O. Hellwig, E. E. Fullerton, G. Denbeaux, J. B. Kortright, and K. Liu, Phys. Rev. B **70**, 224434 (2004).

49  D. A. Gilbert, G. T. Zimanyi, R. K. Dumas, M. Winklhofer, A. Gomez, N. Eibagi, J. L. Vicent, and K. Liu, Sci. Rep. **4**, 4204 (2014).

50  J. A. De Toro, M. Vasilakaki, S. S. Lee, M. S. Andersson, P. S. Normile, N. Yaacoub, P. Murray, E. H. Sánchez, P. Muñiz, D. Peddis, R. Mathieu, K. Liu, J. Geshev, K. N. Trohidou, and J. Nogués, Chem. Mater. **29**, 8258 (2017).

51  E. C. Burks, D. A. Gilbert, P. D. Murray, C. Flores, T. E. Felter, S. Charnvanichborikarn, S. O. Kucheyev, J. D. Colvin, G. Yin, and K. Liu, Nano Lett. **21**, 716−722 (2021).

52  W. Li, X. Xu, T. Gao, T. Harumoto, Y. Nakamura, and J. Shi, J. Phys. D: Appl. Phys **55**, 275004 (2022).

53  M. Tabuchi, M. Takahashi, and F. Kanamaru, J. Alloys Compd. **210**, 143 (1994).

54  R. Yu, X. Chong, Y. Jiang, R. Zhou, W. Yuan, and J. Feng, RSC Adv. **5**, 1620 (2015).

55  A. L. Patterson, Phys. Rev. **56**, 978 (1939).

56  X. Shen, A. Chikamatsu, K. Shigematsu, Y. Hirose, T. Fukumura, and T. Hasegawa, Appl. Phys. Lett. **105** (2014).

57  W. Zhou, C. T. Ma, T. Q. Hartnett, P. V. Balachandran, and S. J. Poon, AIP Adv. **11**, 015334 (2021).

58  A. Foley, J. Corbett, A. Khan, A. L. Richard, D. C. Ingram, A. R. Smith, L. Zhao, J. C. Gallagher, and F. Yang, J. Magn. Magn. Mater. **439**, 236 (2017).

59  D. A. Gilbert, J. W. Liao, L. W. Wang, J. W. Lau, T. J. Klemmer, J. U. Thiele, C. H. Lai, and K. Liu, APL Mater. **2**, 086106 (2014).

[60] J. Nogués and I. K. Schuller, J. Magn. Magn. Mater. **192**, 203 (1999).

[61] S. M. Zhou, K. Liu, and C. L. Chien, Phys. Rev. B **58**, R14717 (1998).

[62] R. L. Stamps, J. Phys. D: Appl. Phys. **33**, R247 (2000).

**Supplementary Material**

**Ionically-Driven Synthesis and Exchange Bias in Mn₄N/MnNₓ Heterostructures**

Zhijie Chen,[1] Christopher J. Jensen,[1] Chen Liu,[2] Xixiang Zhang,[2] and Kai Liu[1,*]

[1]Physics Department, Georgetown University, Washington, DC 20057, USA

[2]King Abdullah University of Science & Technology, Thuwal 23955-6900, Saudi Arabia

## 1. Grazing incidence X-ray diffraction

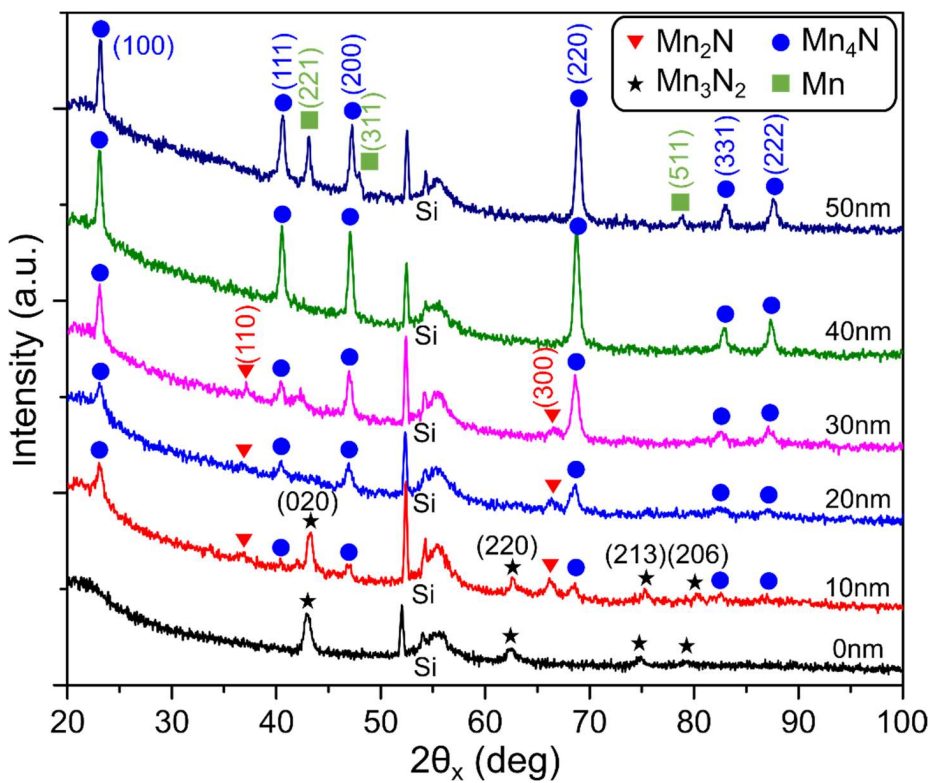

*Figure S1. Grazing incidence X-ray diffraction on samples with different $t_{Mn}$ showing the evolution of Mn₃N₂ (black star), Mn₂N (red triangle), Mn₄N (blue circle), and Mn (green square) phases. $t_{Mn} = 0$ nm is the 20nm Mn₃N₂ seed layer.*

These scans were done on samples with deposited Mn thickness ($t_{Mn}$) from 0 nm to 50 nm. Starting with $t_{Mn} = 0$ nm, which is the 20 nm Mn₃N₂ seed layer (black line), all the peaks (black

Supplementary Material

star) are from the Mn₃N₂ phase. As $t_{Mn}$ increases to 10 nm (red line), Mn₄N peaks (blue circle) emerge because of the reaction between Mn and nitrogen from Mn₃N₂. Mn₂N peaks (red triangle) also show up as Mn₃N₂ loses nitrogen. At $t_{Mn}$ = 20 nm, Mn₃N₂ peaks are all gone after losing too much nitrogen while the Mn₄N peaks grow. Mn₂N peaks, on the other hand, persist until $t_{Mn}$ = 40 nm as the Mn₂N phase loses nitrogen and turns into Mn₄N. In the meantime, Mn₄N peaks grow taller and sharper. At $t_{Mn}$ = 50 nm, there is no nitrogen available for Mn to react with and form Mn₄N. Thus, α-Mn peaks show up. These results are consistent with the interpretation of the 2θ-ω X-ray diffraction shown in Fig. 2(a) of the main text.

## 2. Full range 2θ-ω X-ray diffraction

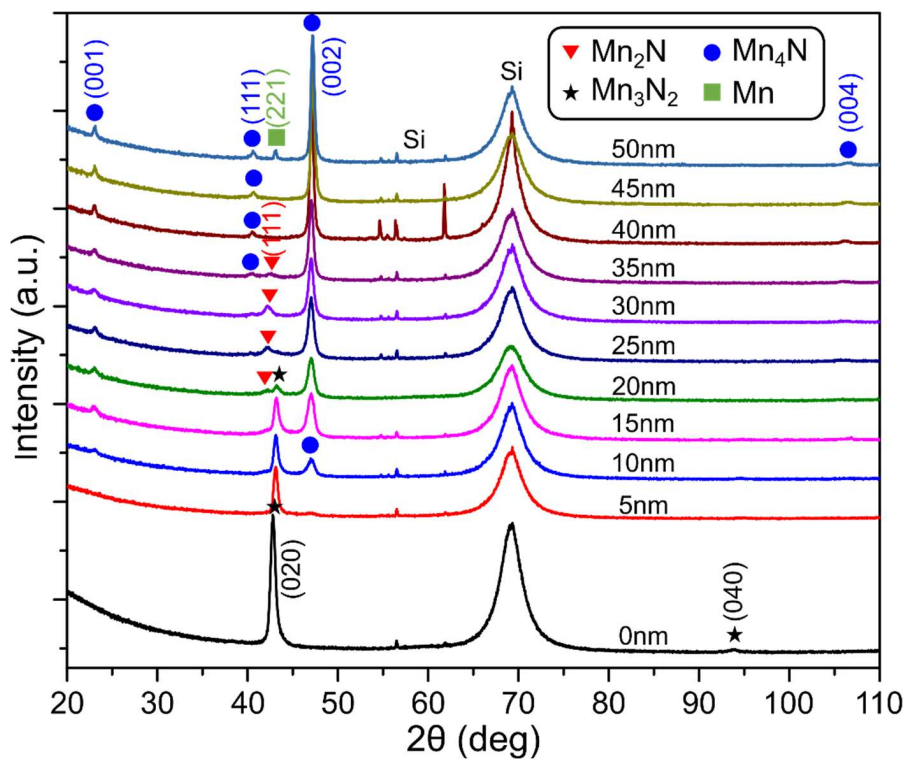

*Figure S2. Full range 2θ-ω X-ray diffraction on samples with different $t_{Mn}$ showing the evolution of Mn₃N₂ (black star), Mn₂N (red triangle), Mn₄N (blue circle), and Mn (green square) phases. $t_{Mn}$ = 0 nm is the 20nm Mn₃N₂ seed layer.*

**Supplementary Material**

These scans are similar to the one shown in Fig. 2(a). These are symmetric scans with a 1° ω offset to suppress the substrate peak. The peaks for the different phases are mainly located between 40° - 50° as shown in Fig. 2(a) of the main text.

**3. Phase evolution**

We have included the expected nitrogen atomic percent and Mn nitride phases identified in each sample in the histogram below (Fig. S3). The expected nitrogen atomic percent is calculated using the nominal $Mn_3N_2$ seed layer thickness and the Mn thickness deposited on top ($t_{Mn}$). The Mn phases in each sample are identified from the XRD peaks in Fig. S1 and S2.

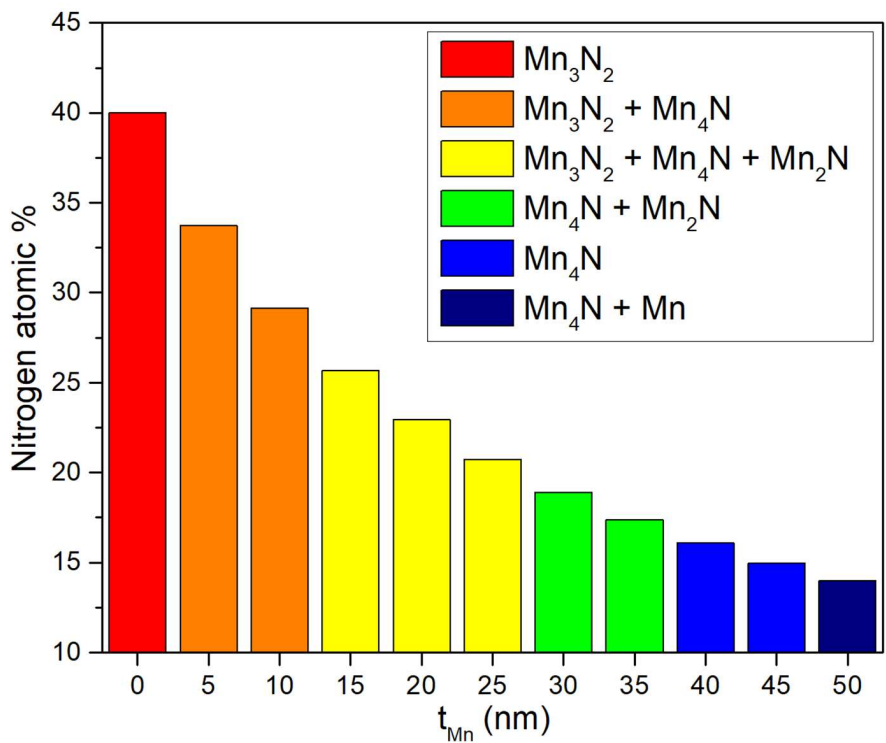

*Figure S3. Histogram showing Mn nitride phases and expected nitrogen atomic percentages in each sample. $t_{Mn}$ is the deposited Mn thickness on top of the 20 nm $Mn_3N_2$. Colors represent different Mn nitride phases.*

**Supplementary Material**

**4. Cross-sectional TEM and EELS line scans**

The cross-sectional TEM lamellas were fabricated using the Helios G4 UX FIB system (Thermo Fisher Scientific) with a $Ga^+$ beam source. Low-energy (2-5 kV) final polishing was employed to minimize the irradiation damage. The composition ratio of Mn and N was determined by EELS line-scan analysis using FEI Titan Themes Cubed G2 300 (Cs Probe) TEM at 300 kV.

EELS line scans are done on samples with $t_{Mn}$ = 0, 20, 40, 50 nm. Those for $t_{Mn}$ = 0 and 40 nm samples are also included in Figs. 2(d) and 2(e) of the main text, and they are shown here for completeness (Fig. S4). The TEM images (left column) indicate the Mn nitride layers are mostly homogenous without distinctive interfaces for all four samples. EELS scans measured across the green lines in the TEM images show that the composition ratio of Mn : N is continuously varying due to the ionic motion of nitrogen within the nitride layers. At $t_{Mn}$ = 0 nm, atomic ratio of Mn:N is 58:42, consistent with the atomic ratio of $Mn_3N_2$. As $t_{Mn}$ increases, nitrogen redistribute within the Mn nitride layers and the Mn:N ratio changes to 66:34 for $t_{Mn}$ = 20 nm and 84:16 for $t_{Mn}$ = 40 nm. This is also consistent with XRD results (Fig. S1 and S2) that indicate $Mn_4N$ is the only nitride phase at $t_{Mn}$ = 40 nm. When $t_{Mn}$ increases to 50 nm, atomic percent ratio further increases to 92:8, likely because of the existence of pure Mn as shown in XRD (Fig. S1 and S2). Note that there is some non-uniformity near the interfacial region between the substrate/capping layer and nitride layers, likely caused by interfacial mixing effect,[1] as nitrogen tends to go into the substrate and capping layer more than Mn.

**Supplementary Material**

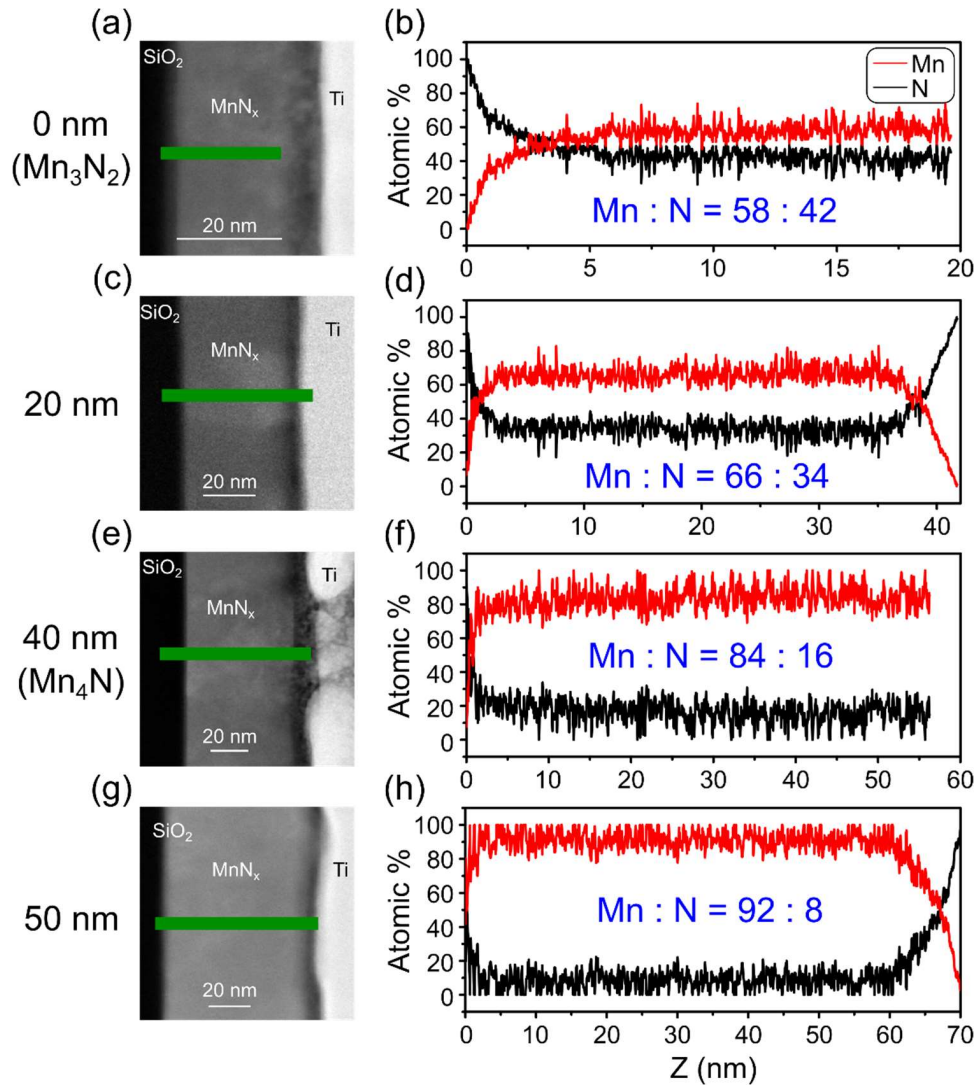

*Figure S4. TEM (left column) and corresponding EELS line scan (right column) along the green line shown in TEM for (a,b) $t_{Mn}$ = 0 nm ($Mn_3N_2$ seed layer), (c,d) $t_{Mn}$ = 20 nm, (e,f) $t_{Mn}$ = 40 nm, and (g,h) $t_{Mn}$ = 50 nm.*

We have also estimated the theoretical Mn/N ratio as $t_{Mn}$ changes. Fig. S5 shows the experimental (EELS) and calculated Mn/N ratio variation as $t_{Mn}$ changes. The theoretical Mn/N ratio has a linear relationship with $t_{Mn}$ since the total nitrogen in the system is fixed, and the experimental ratio largely agrees with the theoretical curve except for the $t_{Mn}$ = 50 nm samples, which has unreacted Mn remaining.

**Supplementary Material**

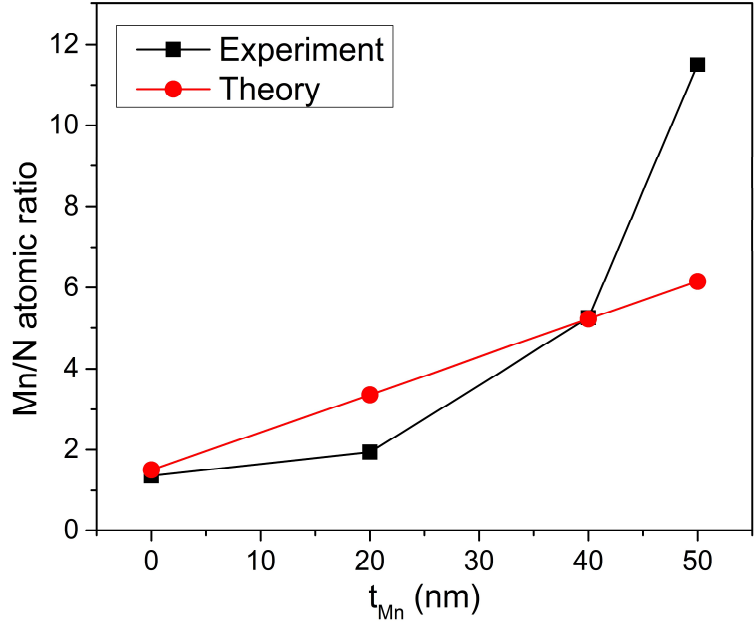

*Figure S5. Experimental (black) and theoretical (red) Mn/N ratio as a function of $t_{Mn}$. Experimental data are determined from EELS results. Theoretical data are determined by fixing the N atoms in the system and change Mn atoms as $t_{Mn}$ changes.*

## 5. Uniaxial magnetic anisotropy calculation

We have also calculated the uniaxial magnetic anisotropy constant ($K_u$) using[2]

$$K_u = K_u^{eff} + \frac{\mu_0}{2} M_S^2, \tag{S1}$$

where $K_u^{eff}$ is the effective anisotropy estimated from the area difference between the IP and OP hysteresis loops and $\frac{\mu_0}{2} M_S^2$ is the thin film demagnetization energy.

## 6. Sample information

The table below lists detailed sample information, including the nominal $Mn_3N_2$ seed layer thickness, deposited Mn thickness ($t_{Mn}$), total thickness determined by X-ray reflectivity, saturation magnetization ($M_s$), $Mn_4N$ in-plane (*a*-axis) and out-of-plane lattice constants (*c*-axis).

**Supplementary Material**

*Table 1. Sample layer thickness information, saturation magnetization, and lattice constants.*

| Sample | Nominal $Mn_3N_2$ seed layer thickness (nm) | Nominal Deposited Mn thickness (nm) | Total thickness (nm) | $M_S$ (RT) (emu/cm$^3$) | $K_u$ (MJ/m$^3$) | $Mn_4N$ $a$-axis (nm) | $Mn_4N$ $c$-axis (nm) |
|---|---|---|---|---|---|---|---|
| #1 | 20 | 0 | --- | --- | --- | --- | --- |
| #2 | 20 | 5 | 27.9 | 25.2 | -0.007 | --- | --- |
| #3 | 20 | 10 | 31.3 | 67.6 | 0.001 | 0.3872 | 0.3859 |
| #4 | 20 | 15 | 36.3 | 78.3 | 0.004 | 0.3873 | 0.3861 |
| #5 | 20 | 20 | 41.4 | 86.3 | 0.002 | 0.3866 | 0.3861 |
| #6 | 20 | 25 | 45.0 | 80.2 | 0.010 | 0.3860 | 0.3862 |
| #7 | 20 | 30 | 50.0 | 67.1 | 0.024 | 0.3860 | 0.3861 |
| #8 | 20 | 35 | 54.4 | 79.1 | 0.030 | 0.3859 | 0.3860 |
| #9 | 20 | 40 | 60.2 | 85.5 | 0.027 | 0.3855 | 0.3856 |
| #10 | 20 | 45 | 65.4 | 71.2 | 0.028 | 0.3846 | 0.3848 |
| #11 | 20 | 50 | 70.5 | 63.5 | 0.020 | 0.3843 | 0.3848 |

**References**

1. C. T. Ma, W. Zhou, B. J. Kirby and S. J. Poon, AIP Adv. **12**, 025117 (2022).

2. T. Hirose, T. Komori, T. Gushi, A. Anzai, K. Toko and T. Suemasu, AIP Adv. **10**, 025117 (2020).